\documentstyle[fleqn]{article}
\begin{document}
\LARGE
\begin{center}
On an Observer-Related Unequivalence Between Spatial Dimensions of a Generic Cremonian Universe
\end{center}
\vspace*{-.2cm}
\Large
\begin{center}
Metod Saniga

\vspace*{.3cm} \small {\it Astronomical Institute, Slovak Academy
of Sciences, 05960 Tatransk\' a Lomnica, Slovak Republic}
\end{center}

\vspace*{-.4cm}
\noindent
\hrulefill

\vspace*{.2cm}
\small
\noindent
{\bf Abstract}\\

\noindent 
Given a generic Cremonian space-time, its three spatial dimensions are shown to exhibit an intriguing, ``two-plus-one"
partition with respect to standard observers.  Such observers are found to form three distinct, disjoint groups based on
which one out of the three dimensions stands away from the other two. These two subject-related properties have, to our knowledge,
no analogue in any of the existing physical theories of space-time. 

\noindent 
\hrulefill

\vspace*{.5cm} \normalsize \noindent 
When confronting a new theory, attention is always paid to those features that make the theory both unrivalled and subject to unambiguous 
falsifiability. 
The theory of Cremonian space-time(s) [1--4] can be no exception in this respect. As for the first aspect, this theory has already been able to
shed a remarkably fresh light on such pressing issues of contemporary physics as the macroscopic dimensionality and signature of the Universe 
[1,2,4], its possible origin and/or evolution [5,7], as well as on a puzzling discrepancy between the physical and psychological/mental concepts of time 
[1,4,6].  The second aspect, its falsifiability, has so far been mentioned in passing only [8] and asks, therefore, for a closer inspection. 

At the current stage of its development, there are very few testable predictions of the theory going beyond the above-mentioned three domains. Yet, 
one of them, although being of a rather subtle nature, stands out as truly fascinating and enormously challenging, for, among other things, it seems to 
undermine the status of two currently most favoured paradigms of natural sciences, viz. reductionism and third-person perspective. The feature
concerned is an {\it un}equal footing on which three ``Cremonian" space dimensions stand with respect to the {\it observer/subject}. Rephrased in
a more explicit way, introducing an observer into our Cremonian space-time breaks the original symmetry by inducing/generating a 
delicate, 2+1 ``splitting-up" in the status of its three spatial dimensions. Mathematically, this fascinating property is intimately connected with
the fact that the intrinsic geometry of a proper conic is identical with that of a projective line and a projectively invariant  property of four 
distinct points of a projective line known as separation [see, e.g., Refs. 9,10].

To begin with, we shall recall that a generic Cremonian space-time [1,4] is an algebraic geometrical configuration that sits in a real 
three-dimensional projective space and comprises three pencils of lines (spatial dimensions) and a single pencil of conics (time). 
The pencils of lines, $\widetilde{{\cal L}}_{\alpha}$ ($\alpha=1,2,3$), are located separately in three distinct planes sharing a line, 
$\widehat{{\cal L}}$, and their respective vertices, $\widehat{B}_{\alpha}$, are assumed not to be collinear, none of them being incident 
with the line $\widehat{{\cal L}}$. The pencil of conics, $\widetilde{{\cal Q}}$, is situated in the plane defined by $\widehat{B}_{\alpha}$ and
its base points are the three vertices and the point, $\widehat{L}$, at which the line $\widehat{{\cal L}}$ meets the plane in question.
In a suitably chosen system of homogeneous coordinates, $\breve{z}_{i}$ ($i=1,2,3,4$), this configuration can analytically be described
as follows:
\begin{equation}
\widetilde{{\cal L}}_{1}(\vartheta_{1}):~~\breve{z}_{2} = 0 = \breve{z}_{3} + \vartheta_{1} \breve{z}_{4},
\end{equation}
\begin{equation}
\widetilde{{\cal L}}_{2}(\vartheta_{2}):~~\breve{z}_{1} = 0 = \breve{z}_{3} + \vartheta_{2} \breve{z}_{4},
\end{equation}
\begin{equation}
\widetilde{{\cal L}}_{3}(\vartheta_{3}):~~\breve{z}_{1} - \breve{z}_{2} = 0 = 2\breve{z}_{3} - \breve{z}_{2} - \breve{z}_{1} + 
\vartheta_{3} \breve{z}_{4},
\end{equation}
and
\begin{equation}
\widetilde{{\cal Q}}(\eta):~~ \breve{z}_{4} = 0 = \breve{z}_{1}(\breve{z}_{3} - \breve{z}_{2}) + \eta \breve{z}_{2}(\breve{z}_{3} -
\breve{z}_{1}),
\end{equation}
where the parameters $\vartheta_{\alpha}, \eta \in \Re^{+} \equiv \Re \cup \{\infty\}$, $\Re$ being the field of real numbers. The equation
of the line $\widehat{{\cal L}}$ obviously reads
\begin{equation}
\widehat{{\cal L}}:~~\breve{z}_{1} = 0 = \breve{z}_{2},
\end{equation}
the three vertices are given by
\begin{equation}
\widehat{B}_{1}:~~\varrho \breve{z}_{i} = (1,0,0,0),
\end{equation}
\begin{equation}
\widehat{B}_{2}:~~\varrho \breve{z}_{i} = (0,1,0,0),
\end{equation}
\begin{equation}
\widehat{B}_{3}:~~\varrho \breve{z}_{i} = (1,1,1,0),
\end{equation}
and the point $\widehat{L}$ as
\begin{equation}
\widehat{L}:~~\varrho \breve{z}_{i} = (0,0,1,0),
\end{equation}
with $\varrho$ being a non-zero proportionality factor.

Central objects of our subsequent reasoning will be proper (non-degenerate) conics of pencil (4). As this pencil, being of the most
general type [1,4,6], contains three distinct composite (degenerate) conics, viz.
\begin{equation}
\widetilde{{\cal Q}}_{0}^{\odot} \equiv \widetilde{{\cal Q}}(\eta=0):~~\breve{z}_{4} = 0 = \breve{z}_{1}(\breve{z}_{3} - \breve{z}_{2}), 
\end{equation}
\begin{equation}
\widetilde{{\cal Q}}_{-1}^{\odot} \equiv \widetilde{{\cal Q}}(\eta=-1):~~\breve{z}_{4} = 0 = \breve{z}_{3}(\breve{z}_{1} - \breve{z}_{2}), 
\end{equation}
and
\begin{equation}
\widetilde{{\cal Q}}_{\infty}^{\odot} \equiv \widetilde{{\cal Q}}(\eta=\infty):~~\breve{z}_{4} = 0 = \breve{z}_{2}(\breve{z}_{3} - \breve{z}_{1}), 
\end{equation}
its proper conics form three distinct, disjoint families $\widetilde{{\cal Q}}^{(\alpha)}$, viz.
\begin{equation}
\widetilde{{\cal Q}}^{(1)}(\eta):~~-1< \eta < 0,
\end{equation}
\begin{equation}
\widetilde{{\cal Q}}^{(2)}(\eta):~~-\infty < \eta < -1,
\end{equation}
and
\begin{equation}
\widetilde{{\cal Q}}^{(3)}(\eta):~~0< \eta < +\infty.
\end{equation}
It is easy to verify that any proper conic of (13)--(15) can be parametrized as
\begin{equation}
\widetilde{{\cal Q}}^{(\alpha)}(\eta):~~\varrho \breve{z}_{i}(\kappa) = (1+\eta (1-\kappa), \kappa, \kappa(1+\eta(1-\kappa)), 0),
\end{equation}
with the parameter $\kappa \in \Re^{+}$. Comparing the last equation with Eqs.\,(6)--(9), and remembering that $\eta \neq -1,0,\infty$,
we find the following correspondence
\begin{equation}
\widehat{B}_{1}:~~\kappa=0,
\end{equation}
\begin{equation}
\widehat{B}_{2}:~~\kappa=1+1/\eta,
\end{equation}
\begin{equation}
\widehat{B}_{3}:~~\kappa=1,
\end{equation}
and 
\begin{equation}
\widehat{L}:~~\kappa=\infty.
\end{equation}
Now, let us have a closer look at the four base points of the pencil of conics. In a general context/setting, three of them,
$\widehat{B}_{\alpha}$, stand on a different footing than the remaining one, $\widehat{L}$; this is fairly obvious from the fact
that $\widehat{B}_{\alpha}$ are vertices of the three space-generating pencils of lines, while $\widehat{L}$ is not. 
What we aim at demonstrating next is that, given a particular proper conic of (4), there even exists a subtle ``splitting up"
in the status of $\widehat{B}_{\alpha}$ themselves.

To furnish this task, we have to introduce a function on a set of four ordered distinct points $P_{i}, i=1,2,3,4,$ of a
real projective line that is a projective invariant, i.e. a function that does not change under collineation and change of
a coordinate system. Taking $x$ and $y$ to be the homogeneous coordinates of the projective line, such a function 
is of the following form [see, e.g., Refs. 9,10]  
\begin{equation}
{\sf CR}\{ P_{1}, P_{2};  P_{3}, P_{4}\}  \equiv \frac{(x_{1}y_{3} - x_{3}y_{1})(x_{2}y_{4} - x_{4}y_{2})}
{(x_{1}y_{4} - x_{4}y_{1})(x_{2}y_{3} - x_{3}y_{2})},
\end{equation}
where $x_{i}, y_{i}$ are the coordinates of the point $P_{i}$. If we switch to a more convenient affine parameter $\kappa \equiv x/y \in \Re^{+}$,
this equation can be rewritten as
\begin{equation}
{\sf CR}\{ P_{1}, P_{2};  P_{3}, P_{4}\}  = \frac{(\kappa_{1} - \kappa_{3})(\kappa_{2} - \kappa_{4})}
{(\kappa_{1} - \kappa_{4})(\kappa_{2} - \kappa_{3})}.
\end{equation}
The function ${\sf CR}\{ P_{1}, P_{2};  P_{3}, P_{4}\}$ is called the {\it cross ratio} of the four points in question and, as easily
verified, satisfies the following symmetry relations:
\begin{equation}
{\sf CR}\{ P_{1}, P_{2};  P_{3}, P_{4}\} = {\sf CR}\{ P_{2}, P_{1};  P_{4}, P_{3}\} = {\sf CR}\{ P_{3}, P_{4};  P_{1}, P_{2}\},
\end{equation}
\begin{equation}
{\sf CR}\{ P_{2}, P_{1};  P_{3}, P_{4}\} = 1/{\sf CR}\{ P_{1}, P_{2};  P_{3}, P_{4}\},
\end{equation}
\begin{equation}
{\sf CR}\{ P_{1}, P_{3};  P_{2}, P_{4}\} = 1 - {\sf CR}\{ P_{1}, P_{2};  P_{3}, P_{4}\}.
\end{equation}
Next, let us view the four points as forming two pairs, say $P_{1}, P_{2}$ and $P_{3}, P_{4}$. As the points are assumed to be distinct, we
can introduce a very important concept of {\it separation}. Namely, the pair $P_{1}, P_{2}$ is said to separate, resp. not to separate, the
pair $P_{3}, P_{4}$  (denoted $P_{1}, P_{2} \| P_{3}, P_{4}$, resp. $P_{1}, P_{2} \sharp P_{3}, P_{4}$) according as 
${\sf CR}\{ P_{1}, P_{2};  P_{3}, P_{4}\}$ is negative resp. positive. It is clear that separation is a projective invariant, endowed -- as it easily
follows from Eqs.\,(23)--(25) -- with the following properties: $a$) if  $P_{1}, P_{2} \| P_{3}, P_{4}$, then $P_{1}, P_{2} \| P_{4}, P_{3}$
and $P_{3}, P_{4} \| P_{1}, P_{2}$; $b$) if  $P_{1}, P_{2} \| P_{3}, P_{4}$, then $P_{1}, P_{3} \sharp P_{4}, P_{2}$
and $P_{1}, P_{4} \sharp P_{2}, P_{3}$; and $c$) there is one and only one way the four points can be divided into two mutually separating
pairs.

We have already mentioned that a proper conic is intrinsically isomorphic to a projective line. So, employing this cross-ratio machinery, we 
are now ready to analyze the separation properties of the four base points of pencil (4) for
each proper conic located in the latter. Thus, combining Eqs.\,(17)--(20) and Eq.\,(22), and recalling that $\eta \neq -1,0,\infty$,
we get
\footnote{To be completely rigorous, in Eqs\,(26)--(28) we should take the limits as $\kappa \rightarrow \infty$ for $\widehat{L}$, but the results
will be the same.}
\begin{equation}
{\sf CR}\{ \widehat{B}_{2}, \widehat{B}_{3};  \widehat{B}_{1}, \widehat{L}\} = 
\frac{(1 + 1/\eta  - 0)(1 - \infty)}{(1+ 1/\eta - \infty)(1 - 0)} = \frac{\eta + 1}{\eta},
\end{equation}
\begin{equation}
{\sf CR}\{ \widehat{B}_{1}, \widehat{B}_{3};  \widehat{B}_{2}, \widehat{L}\} = 
\frac{(0 - 1 - 1/\eta)(1 - \infty)}{(0 - \infty)(1 - 1 - 1/\eta)} = \frac{-1 - 1/\eta}{-1/\eta} =  \eta +1,
\end{equation}
\begin{equation}
{\sf CR}\{ \widehat{B}_{1}, \widehat{B}_{2};  \widehat{B}_{3}, \widehat{L}\} = 
\frac{(0 - 1)(1 + 1/\eta - \infty)}{(0 - \infty)(1 + 1/\eta - 1)} = \frac{-1}{1/\eta} =  - \eta.
\end{equation}
If  one takes into account the above-listed properties of separation, then the first of these equations tells us that 
$\widehat{B}_{2}, \widehat{B}_{3} \|  \widehat{B}_{1}, \widehat{L}$ only for proper conics of  the first family (Eq.\,(13)),
the second equation implies that $\widehat{B}_{1}, \widehat{B}_{3} \|  \widehat{B}_{2}, \widehat{L}$ only for proper
conics of the second family (Eq.\,(14)), and, finally, the third equation says that $\widehat{B}_{1}, \widehat{B}_{2} \|  \widehat{B}_{3}, \widehat{L}$
only for proper conics of the third family (Eq.\,(15)). This means that, given the separation property, whatever proper conic of pencil (4)
is chosen, there is always one of the three points $\widehat{B}_{\alpha}$, and, so, one of the three corresponding space dimensions,
that stands slightly aside from the other two; which this point (spatial dimension) is depends on the family the conic in question belongs to.  
 
The final step of our analysis is to show that there exists a natural way of selecting a particular conic of pencil (4). And, indeed, this can simply
be done in terms of introducing an observer/subject. For every observer in our Cremonian universe is represented by a projective line
and the latter, when in a ``standard" position, cuts the carrier plane of the pencil of conics in a single point that falls on a unique proper
conic [1--4,6]. So for every ``standard" observer, i.e. the observer whose representing line hits the proper conic of pencil (4), space exhibits an intriguing
two-to-one partition among its dimensions; moreover, these observers obviously form three different, disjoint groups according as which of
the three spatial dimensions stands apart from the other two, or, what amounts to the same, which of the three families of proper conics
(Eqs.\,(13)--(15)) hosts the conic selected. And as this unique conic represents nothing but the moment of the present, the ``now" for this particular 
observer [4,6], one can equally claim that the observers sharing the same present moment, as well as all the observers whose ``nows" belong to 
the same family, will experience the same 2+1 space splitting algebra. 

So, if the structure of the deeper levels of the Universe is Cremonian (as we firmly believe), then every standard observer should face, or experience, a 
sort of  ``dissociation" of space, whose character is conditioned by the very existence of the observer and shared by all the standard observers 
in the corresponding group. 
At present, it is rather difficult to -- even conceptually -- envisage a kind of an experiment that would be able to detect this subject-based, 2+1 
breakup of space, which, to the best of our knowledge, has no proper counterpart within the physical theories of space-time. One should, however, not 
be surprised if first verifications come from psychology and neurosciences rather than physics; after all, it is non-ordinary
forms of mental space-time(s) [6] where our Cremonian approach has been found to perform so well.   
\\
\\
\noindent
{\bf Acknowledgements}\\ 
I am very grateful to Avshalom Elitzur and Lawrence Schulman for a host of interesting and motivating comments/questions. 
This work was supported in part by a
2001--2002 NATO/FNRS Advanced Research Fellowship and the 2001--2003 CNR-SAV joint research project ``The Subjective Time
and its Underlying Mathematical Structure."

\end{document}